\input epsf
\documentclass[nohyper,letterpaper,11pt]{JHEP}

\title{On the Masses of the Leptons, Bosons and Quarks}

\author{Roger C. Millikan\\
Department of Chemistry and Biochemistry\\
University of California Santa Barbara CA 93106-9510\\
millikan@chem.ucsb.edu}

\author {Dennis C. Richman\\
Northrop Grumman Corporation\\
Los Angeles, California  90067-2199\\
drichman@ryanaero.com}

\abstract{
A previously developed lepton mass equation is extended to include the 
massive bosons and quarks of all three generations. The particles are 
modeled as closed, string-like, light front solitons whose key quantum 
numbers are their node number n and their winding number $\omega$. The 
simplicity and form of the mass equation suggests that the dominant mass 
effect comes from how the particles embed themselves in spacetime.}

\preprint{Version 6/12/2001}
\begin{document}

\section{Introduction} 
From a postulated model system for the elementary particles \cite{Mill}, we have 
found a simple mass equation that relates the observed rest mass of the 
massive leptons to the mass of the W boson. The relationship in logarithmic 
form is:
\begin{equation} \ln m  = 
-\frac{1}{2\pi\alpha}\left(\frac{1}{2}\right)^{\omega} + \ln \left(
\frac{m_W}{\pi}\right)
\end{equation}
where $m$ is the lepton mass in MeV, $\alpha$ is the fine structure constant (1/137), 
$\omega$  is the winding number of the particle gauge field and equals 1, 2, or 3 for 
the electron, muon, and tau respectively;  $m_W$ = 80,430 MeV (the observed 
mass of the W boson). A plot of this relationship is shown in Fig. 1 where 
the points are the experimental rest masses for the three generations of 
massive leptons and the line is given by Eqn. (1.1). While the fit shown in  
Fig. 1 looks good, one wonders if it is really significant or perhaps just 
coincidence. Having only three points on the line is worrisome. We address 
this worry here by showing how Eqn. (1.1) can be extended to include the 
massive gauge bosons and the quarks. The resulting equation, when 
interpreted physically in model terms, leads to surprising conclusions 
concerning the origin of particle mass, and why the fermion masses 
(excepting the top quark) are so small compared to the boson masses.
\FIGURE{
\epsfbox{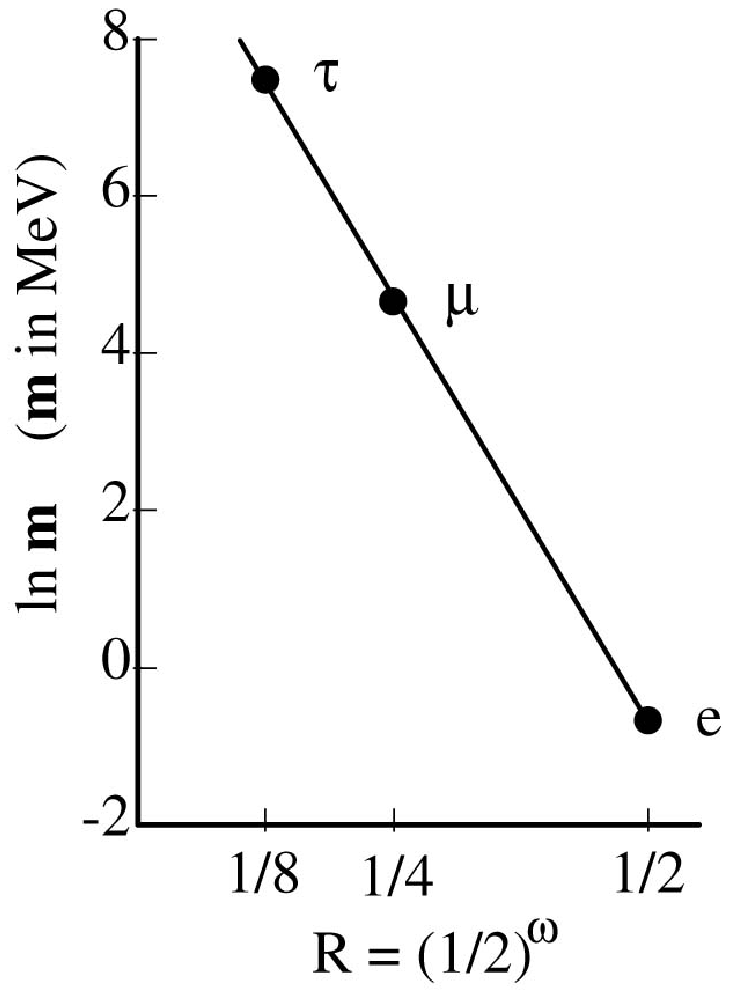}
\caption{Lepton rest masses plotted as ln$(m)$ against a factor R which
accounts for the effect upon mass of the winding number $\omega$ in each 
generation. For the electron, muon, and tau the values of $\omega$ are
1, 2, and 3 respectively. The line comes from Eqn. 1.1}
}

Our model system for the elementary particles [1] will be summarized 
briefly in order to define the quantum numbers and features important in 
determining their masses. Then we discuss how the mass equation can be 
extended to include the bosons and quarks. Lastly we consider the physical 
interpretation and the implications of the mass equation.
\section{Basic features of the particle \\
models.}  
We postulate that:

$\bullet$ First generation massive particles are light-front solitons made of 
transverse harmonic oscillator excitations of U(1) gauge fields 
propagating in a circle at v = c. 

$\bullet$ For once around, the leptons have one node (n = 1), the bosons have two nodes  
\mbox{(n = 2),} and the quarks have three nodes (n = 3).

$\bullet$ The gauge fields have a polarization that twists about the propagation 
path by $\pi$ radians as the front propagates from one node to the next. This 
is required for the wave function to consistently meet the cyclic boundary conditions.

$\bullet$ The fermions (leptons and quarks) are spinor particles; their fields must 
go twice around for a complete cycle of the U(1) field. The bosons are vector 
particles; their fields only go once around for a complete U(1) cycle.

$\bullet$ The twist direction may or may not reverse at a node. This distinguishes 
one particle from another. For example, the $W^-$ has the twist pattern L-L 
for one cycle, while the Zû has a twist pattern of L-R or R-L. So the W 
has a total twist of 2$\pi$ radians per cycle while the Zû has a total twist of 
zero. This is the source of the mass splitting between the W (80.4 GeV) 
and the Z (91.2 GeV).

$\bullet$ First generation particles have a writhe = 0. That is, they have planar 
propagation paths. Second generation particles have writhe = 1, and third 
generation particles have writhe = 2. But since the more physical 
``winding number" equals (writhe + 1), we use winding numbers $\omega$ = 
1,2,3 as the generation quantum number. 

\FIGURE{
\epsfbox{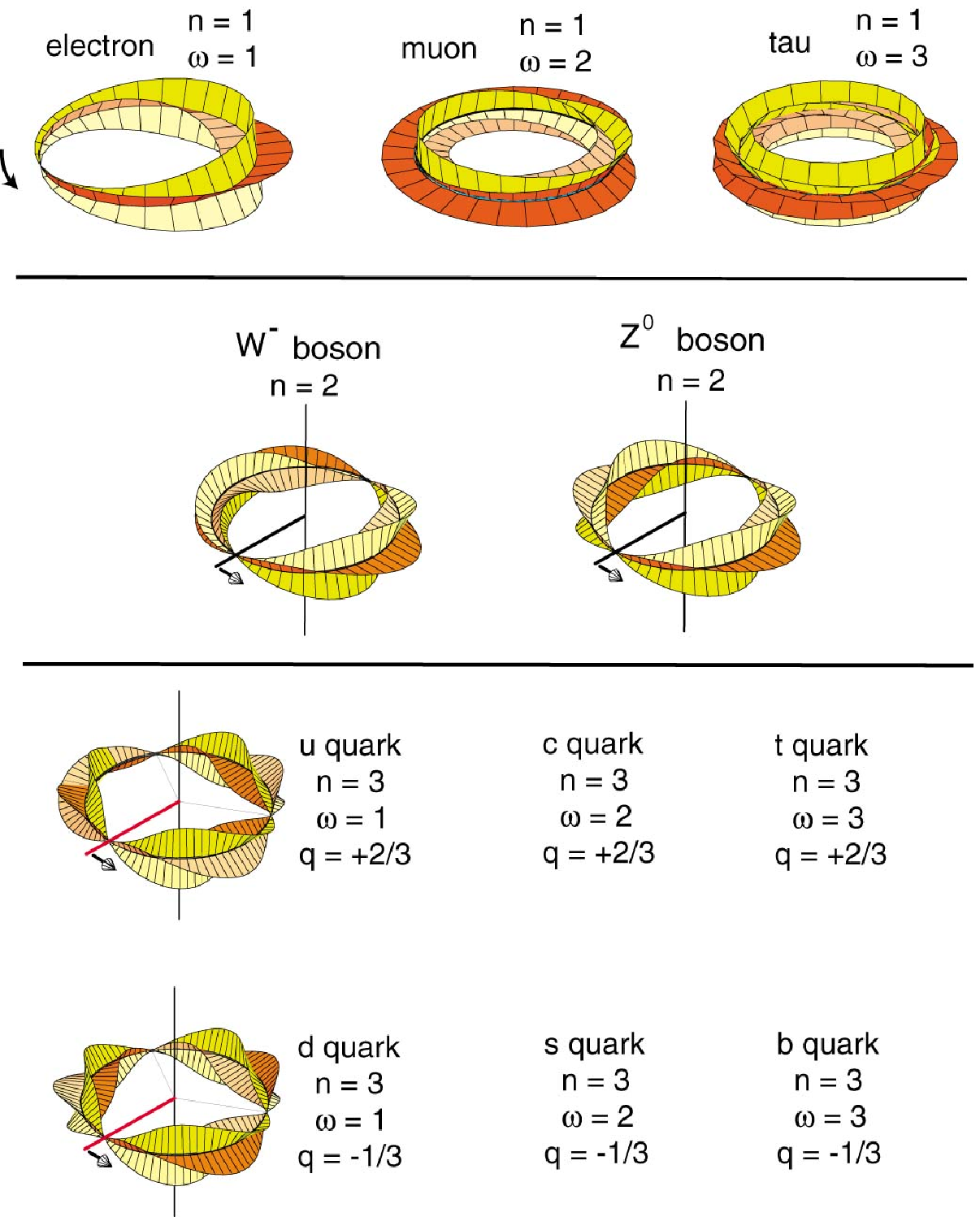}
\caption{Depictions of light front models of the massive leptons, bosons
and quarks. The quantum number \textbf{n} denotes the number of nodes. The winding number
is given by $\omega$. Second and third generation quarks are not shown,
but their quantum numbers are listed.}
}

These features are embodied in the model depictions of Fig. 2. 
At the top of 
the figure the one-node lepton configurations are shown for the three 
generations. Intense orange and light orange colors (dark gray and light gray 
in monochrome) represent the minus and plus parts of the electric field 
respectively. The central ring is an event horizon making the interior fields 
unobservable. Hence a distant observer sees these as negatively charged 
particles when time averaged over many cycles. Bright yellow (lighter gray) 
and light yellow (very light gray) denote the north and south parts of the 
magnetic field. Because the twist inversion of the geometry offsets the cyclic 
inversion of the field, the north half of the field is always atop the ring, and 
the south half is always below. Thus a distant observer sees these particles as 
magnetic dipoles. The central part of Fig. 2 shows the two-node charged and 
neutral boson configurations. It is an experimental fact that the bosons do 
not show generational behavior. A possible reason for this will emerge from 
an interpretation of the mass relationship we will develop. At the bottom of 
Fig. 2 the models for the first generation u and d quarks are shown, and the 
quantum numbers for all the quarks are given. We do not show the field 
configurations for the higher generation quarks. With the higher winding 
numbers, the quark fields become so complex that the depictions are not 
very useful. The quark models have three degenerate forms that 
accommodate the color degree of freedom, but we show only one color 
form.
\section{Known and estimated particle data}

\TABLE{
Table 1. Presently accepted mass values for the elementary particles\\
and their model quantum numbers. \\
\begin{tabular}
[b]{|c|c|c|c|}
\hline \textbf{Leptons} &\textbf{Mass in MeV} &\textbf{Node number} &\textbf{Winding number}\\
& \textbf{from Ref 2} & \textbf{n} &  $\omega$ \\ 

\hline \textbf{e} & 0.511 & 1 & 1 \\ 
\hline $\mathbf{\mu}$ & 105.7 & 1 & 2 \\ 
\hline $\mathbf{\tau}$ & 1777 & 1 & 3 \\
\hline \textbf{Bosons}\\ 
\hline \textbf{W} & 80,430 & 2 & - \\ 
\hline \textbf{Z} & 91,188 & 2 & - \\ 
\hline \textbf{Quarks}\\
\hline \textbf{u} & 1.5 - 5 & 3 & 1 \\ 
\hline \textbf{d} & 3 - 9 & 3 & 1 \\
\hline \textbf{c} & 1150 - 1350 & 3 & 2 \\
\hline \textbf{s} & 60 - 170 & 3 & 2 \\
\hline \textbf{t} & 174,300 & 3 & 3 \\
\hline \textbf{b} & 4000 - 4400 & 3 & 3 \\
\hline
\end{tabular}
}
\TABLE{

Table 2. Mass values for the unsplit parents of the quark doublets,\\
along with their quantum numbers. \\ 
\begin{tabular}
[b]{|c|c|c|c|}
\hline \textbf{Parent level} &\textbf{Average Mass} &\textbf{Node number} &\textbf{Winding number}\\
\textbf{designation} & \textbf{in MeV} & \textbf{n} &  $\omega$ \\ 
\hline \textbf{ud} & 4.6 & 3 & 1 \\ 
\hline \textbf{cs} & 683 & 3 & 2 \\ 
\hline \textbf{tb} & 89,250 & 3 & 3 \\
\hline
\end{tabular}
}
\FIGURE{
\epsfbox{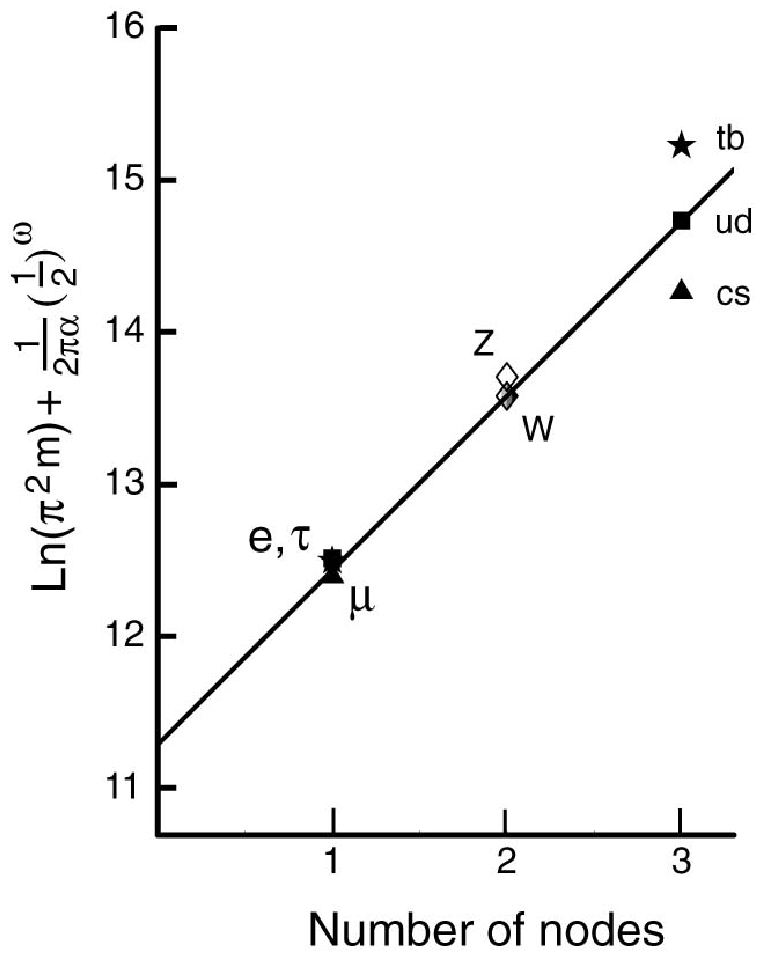}
\caption{Mass plot for the elementary particles as a function of the number
of gauge field nodes for once around the ring. The effect of winding number
is taken out by the way the ordinate is plotted. The line has slope = ln $\pi$
and intercept = ln($m_W$).}
}
The masses of the leptons plotted in Fig. 1 have been experimentally 
measured and are very well known \cite{PDG}. Since the quarks have never been 
isolated, their masses must be deduced indirectly from many measurements 
as interpreted by QCD theory. 
For the lighter quarks, only a range of values 
can be specified. Collected in Table 1
are these mass values along with the 
relevant quantum numbers for the different particles. Where a range of mass 
is given, we use the average value. Also, in each generation of quark, the 
doublet of masses appears to be split from some parent level. At this stage 
we do not understand the splitting interaction, so we simplify by assuming 
that the observed mass levels come from a parent level half way in between. 
The masses for these unsplit levels are given in Table 2.
For the bosons, no entry is given in the winding number column. This is 
because the bosons do not show generational behavior.
\section {Extending the formula to \\
include the bosons and quarks}
	For particles with more than one node, we expect a contribution to the 
mass from ``nodal" excitations. In order to see what this might be we have 
manipulated Eqn. 1.1, moving everything to the left hand side except the node 
dependence and plotted that against the number of nodes as in Fig. 3. 
For the W and Z bosons the generational term has been eliminated by taking 
the limit as $\omega$ goes to infinity (This has a physical basis that will be 
discussed later). The linear relationship with a slope of log $\pi$ as seen in Fig. 
3 tells us that each node represents a factor of $\pi$ contribution to the particle 
mass equation. Knowing this lets us write an extended mass equation:
\begin{equation} 
\ln \left(\frac{m\pi^2}{\pi^n}\right) = 
-\frac{1}{2\pi\alpha}\left(\frac{1}{2}\right)^{\omega} + \ln \left(m_W\right)
\end{equation}
where the symbols are as in Eqn. 1.1 and n is the particle node number taken 
from Tables 1 and 2. For the bosons, we set  $\omega = \infty$ which removes the 
first term on the rhs of Eqn. (4.1) for them. This equation is plotted as the line 
in Fig. 4, along with points for all the elementary particles using the data 
from Tables 1 and 2.
\FIGURE{
\epsfbox{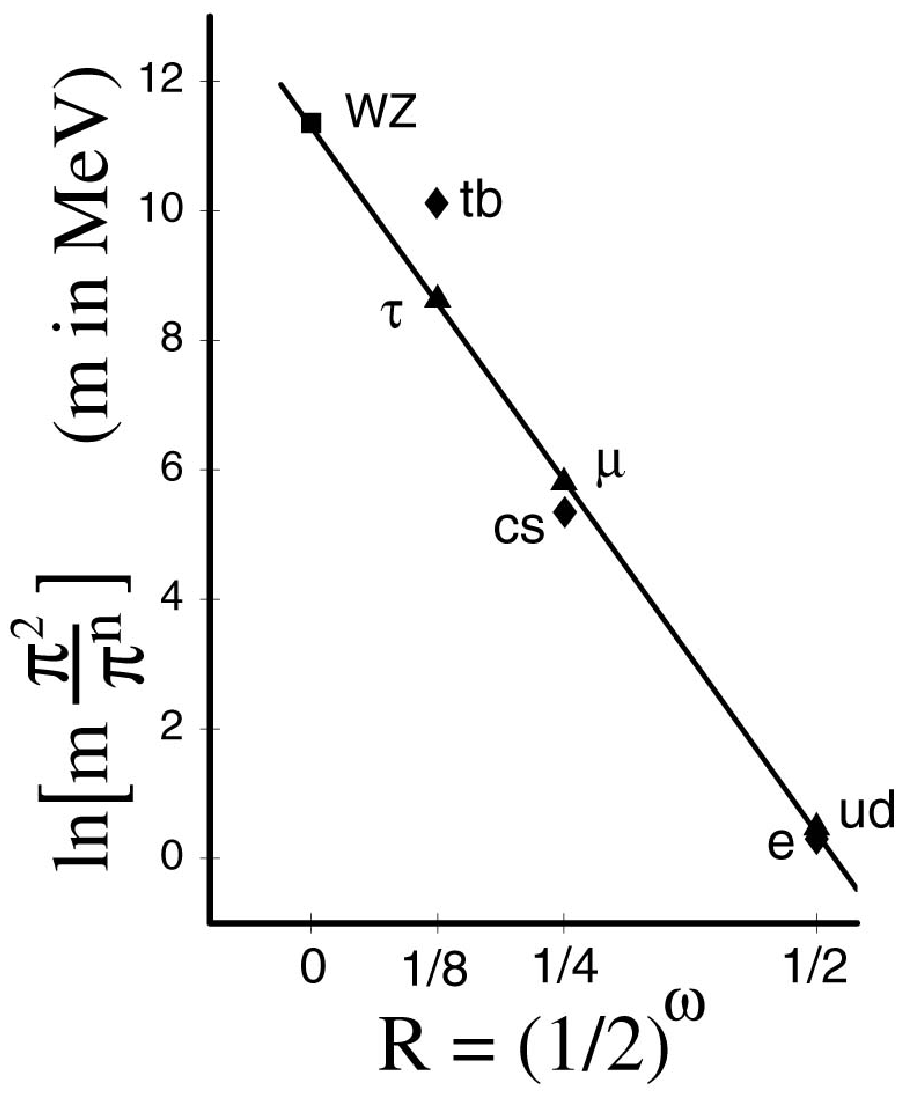}
\caption{Plot of lepton and average quark masses for each generation
against the winding number factor. Also shown is the point for the average
of W and Z boson masses for \textbf{R} = 0. The node number \textbf{n} is 1, 2, 3 for
the leptons, bosons, and quarks respectively. The winding number \textbf{$\omega$}
is 1, 2, 3 for the first, second, and third generations of the fermions. The line
has slope = -1/(2$\pi\alpha$) and intercept = ln $m_W$).}
} 

As expected from Fig. 1, the experimentally known masses of the leptons 
and bosons fit the line given by Eqn. 4.1 quite well. The currently accepted 
masses of the quarks [2], with the doublets combined as described show 
more scatter about the line. Yet they are reasonably well represented given 
the approximations made. Thus we take this mass relationship as a valid 
equation for the masses of the leptons, bosons and quarks. Assuming that, 
the natural question is: ``What features and interactions of the particles could 
give rise to such an equation?"
\section {Physical interpretation of the mass equation}
	As a start, we exponentiate Eqn. 4.1, and rearrange it into a form that is 
easier to interpret:
\begin{equation} 
m = m_W\bullet\frac{1}{\pi^2}\bullet\pi^n\bullet
\exp\left[\frac{-1}{2\pi\alpha}\left(\frac{1}{2}\right)^\omega\right]
\end{equation}
The explicit multiplication dots are given to emphasize an important 
aspect; the different contributions to the mass are multiplicative, not 
additive.  This can be understood if Eqn. 5.1 is a partition function equation. In 
statistical mechanics the total partition function for a system having 
independent parts is the product of the partition functions for the parts. This 
also shows up in topological quantum field theories where the vacuum 
expectation value of the field is given by the product of Wilson loop 
holonomies.\cite{Witt} Indeed, in those theories the (vev) is called a partition 
function. From this point of view, we can think of Eqn. 5.1 as giving the 
partitioning of the mass-energy of the particle among the degrees of freedom 
open to it.
The $m_W$ factor sets the mass scale for all the particles. The $\pi^{-2}$ factor 
should be considered a normalization factor. Since the W boson has two 
nodes (n=2), this factor cancels the next factor when the equation is applied 
to the W boson. The $\pi^n$ factor  accounts for the level of excitation of the 
field harmonic oscillators in terms of the node number. The last factor, the 
exponential, is different in several respects. It may be thought of as an 
``embedding factor", giving the fraction of the intrinsic particle mass that is 
observable to a real observer far away. This is the factor that 
distinguishes the masses of the different generations according to their 
winding numbers. For the bosons we allow the winding number to go to 
infinity, and this factor becomes unity. This sets the fermions apart from the 
bosons in a way discussed below.

One of the surprising features of Eqn. 5.1 is its simplicity. When one 
considers the mass of an elementary particle, one expects contributions from 
many sources. For charged particles there is the energy stored in the 
electromagnetic field. There was a time when the entire mass of the electron 
was thought to reside in its electromagnetic field. We now know that this 
contribution to the mass is very small compared to that arising from other 
sources. Then there are radiative corrections and possible quantum 
corrections that may contribute to particle masses. They seem not to be 
included in Eqn. 5.1. Lastly, contributions from internal excitations should 
contribute to the particle mass. These are included as the $\pi^n$ factor in our 
equation. But comparing the relative contributions, it is clear that this one is 
minor to that specified by the $m_W$ factor. Here is a clue as to why this mass 
equation is so simple. There is one overwhelming and dominant effect that 
makes all the other contributions negligible. What is this dominant effect? 
We believe it is the need to provide energy to embed the rotating particle in 
the surrounding spacetime.  Equation 5.1 is a ``winding number" equation in 
the following sense. According to our model, these particles are quantum 
black holes of Kerr geometry that are rotating at v = c. From studies of Kerr 
black holes applied to rotating stars, it is understood that near the event 
horizon, the whole of spacetime is set into rotation. It is like a vortex in fluid 
mechanics in which the surrounding fluid has a rotating velocity field 
surrounding the vortex core. When a stable particle is formed, it must have 
enough energy to set up the vortex field in the surrounding spacetime. This 
is the dominant energy term for these particles that Equation 5.1 describes, and 
why $m_W$ is so large. Yet somehow the fermions seem to evade the bulk of 
this energy cost. How? It is because they are spinor particles. The rotation of 
spin 1/2 particles is very special. When a spin 1/2 particle is rotated about 
any axis for $2\pi$ radians, it winds up its connections to the surroundings just 
like a boson. But when the spinor is rotated another $2\pi$ radians, it can 
unwind the connection to the surroundings by suitable writhing. This 
possibility was known to Dirac, and is well described in the book of Misner, 
Thorne, and Wheeler\cite{MTW}. Applied to our particle models we have this 
picture. For the bosons, as time passes, their fields continuously wind 
around, frame-dragging nearby spacetime with them. This is the origin of the 
infinite winding number for the bosons. For the fermion particles, as their 
fields rotate around, they cyclically wind the surrounding spacetime on the 
first  2$\pi$ revolution and then unwind it on the second 2$\pi$ revolution. The 
result is that the fermions have to spend very much less energy embedding 
themselves into their surroundings than do the bosons.
One can see this effect at work in the exponential factor of Eqn. 5.1. 
The internal winding number $\omega$ specifies how many times the field goes 
around before the node is reached (the half cycle point) where the spinor can 
begin unwinding its connection. This translates into the fraction of the 
energy for infinite winding that the particle must have.
If this picture is correct, it implies that all the massive fermions should 
be much less massive than any of the massive bosons. This is in fact true 
excepting the top quark, as can be seen by looking at Table 1. From our mass 
correlation, it appears that the top quark is a special case and has a higher 
mass than expected. That is the why the tb quark point lies above the line in 
both Fig. 3 and 4. One possible reason for this is that some new effect not 
included in our considerations (Supersymmetry mixing?) is beginning to show
up at the highest 
winding numbers and node numbers. Perhaps the same effect keeps fourth 
and higher generation families from having observable particles.
\section{Discussion}
The particle mass equation 5.1 describes mainly how the various 
particles embed themselves into spacetime. The mass effects connected with 
that dominate all the other mass effects that one usually thinks about. It is 
surprising that the mass equation contains no adjustable parameters if one 
considers $m_W$ and $\alpha$ as constants of nature. Clearly, much additional work 
needs to be done to fully appreciate what this mass equation is telling us.


\begin{thebibliography}{4}

\bibitem{Mill} R. C. Millikan, \emph{Light Front Models for the Leptons,
Bosons, and Quarks}, previous paper, 
\href{http://www.chem.ucsb.edu/millikan/Paper.html}{http://www.chem.ucsb.edu/millikan/Paper.html},
hep-th/0106098.


\bibitem{PDG} Particle Data Group, \emph{Review of Particle Physics}, {\it European Phys. J.(C)}, 
{\bf 15}, (2000) 1-878.

\bibitem{Witt} E. Witten, \emph{Quantum Field Theory and the Jones Polynomial},
{\it Comm. Math. Phys. } {\bf 121} (1989) 351.

\bibitem{MTW} C. W. Misner, K. S. Thorne, and J. A. Wheeler, \emph{Gravitation} (W. H. 
Freeman and Company, San Francisco 1973) p. 1148ff. 
\end{thebibliography}
\end{document}